\documentclass[prd,onecolumn]{revtex4}
\usepackage{dcolumn}
\usepackage{multirow}
\usepackage{graphicx}
\usepackage{amssymb}
\usepackage{bm}
\usepackage{hyperref}
\usepackage{epstopdf}
\usepackage{color}
\usepackage{mathrsfs}
\usepackage{amsmath,amssymb,amsthm}
\begin{document}
\title{Geometric dark energy traversable wormholes constrained by astrophysical observations}
\author{Deng Wang}
\email{Cstar@mail.nankai.edu.cn}
\affiliation{Theoretical Physics Division, Chern Institute of Mathematics, Nankai University,
Tianjin 300071, China}
\author{Xin-he Meng}
\email{xhm@nankai.edu.cn}
\affiliation{{Department of Physics, Nankai University, Tianjin 300071, P.R.China}\\
{State Key Lab of Theoretical Physics,
Institute of Theoretical Physics, CAS, Beijing 100080, P.R.China}}
\begin{abstract}
In this letter, we introduce the astrophysical observations into the wormhole research, which is not meant to general parameters constraints for the dark energy models, in order to understand more about in which stage of the universe evolutions wormholes may exist through the investigation of the evolution behavior of the cosmic equation of state parameter. As a concrete instance, we investigate the Ricci dark energy (RDE) traversable wormholes constrained by astrophysical data-sets. Particularly, we can discover from Fig. \ref{fig5} of the present work, when the effective equation of state parameter $\omega_X<-1$, namely, the Null Energy conditions (NEC) is violated clearly, the wormholes will appear (open). Subsequently, six specific solutions of static and spherically symmetric traversable wormhole supported by the RDE are obtained. Except for the case of constant redshift function, in which the solution is not only asymptotically flat but also traversable, the remaining five solutions are all not asymptotically flat, therefore, the exotic matter from the RDE fluids is spatially distributed in the vicinity of the throat. Furthermore, we analyze the physical characteristics and properties of the RDE traversable wormholes. It is worth noting that, through the astrophysical observations, we get constraints on the parameters of RDE model, explore the type of exotic RDE fluids in different stages of the universe changing, limit the number of available models for wormhole research, reduce the number of the wormholes corresponding to different parameters for RDE model and provide a more apparent picture for wormhole investigations from the new perspective of observational cosmology background.
\end{abstract}
\maketitle
\section{Introduction}
Modern astronomical observations with increasing evidence (such as high redshift Type Ia supernovae (SNe Ia), matter power spectra, observational Hubble parameter data (OHD), cosmic microwave background radiation (CMBR), etc.) have strongly suggested that the universe is undergoing an accelerated phase at present\cite{Riess,0,1,2,3}. To explain the accelerated mechanism, cosmologists have proposed a new negative pressure fluid named dark energy. The simplest candidate of dark energy is the so-called cosmological constant, namely, $\Lambda$CDM model, which is proved to be very successful in describing many aspects of the observed universe. For instance, the spectrum of anisotropies of the CMBR, the large scale structure of matter distribution at linear level (LSS), and the expansion phenomena are very well described by the standard cosmological model. However, this model has faced two fatal problems, namely, the `` fine-tuning '' problem and the `` coincidence '' problem \cite{Weinberg}. The former indicates that theoretical estimates for the vacuum density are many orders of magnitude larger than its observed value, i.e., the famous 120-orders-of-magnitude discrepancy that makes the vacuum explanation suspicious, while the latter implies that why the dark energy and dark matter are at the same order today since the energy densities of them are so different during the evolution of the universe. In addition, a positive cosmological constant is inconsistent with perturbed string theory \cite{Witten}. Therefore, the realistic nature of dark energy should not be simply the cosmological constant $\Lambda$ (interpreting it as quantum vacuum). In recent years, to alleviate or even solve these two problems, cosmologists have proposed a variety of dark energy models, partly as follows

$\bullet$ Exotic equation of state: a linear equation of state \cite{A}, Van der Waals equation of state \cite{B,C}, Chaplygin gas \cite{Alexzander,Xu,D}, generalized Chaplygin gas \cite{E,F}, modified Chaplygin gas \cite{G,H}, superfluid Chaplygin gas \cite{I,J,K}, inhomogeneous equation of state \cite{K0}, barotropic fluid model \cite{K1}, Cardassian model \cite{L,M,N,O}.

$\bullet$ Viscosity: bulk viscosity in the isotropic space, bulk and shear viscosity in the anisotropic space \cite{rm,rm0,rm1,rm2,rm3,rm4}. It is worth noting that the perfect fluid that occurs in many papers is just an approximation of the universe media. Nowadays, all the observations indicate that the universe media is not an idealized fluid and the viscosity is investigated in the evolution of the universe.

$\bullet$ Holographic Principle: holographic dark energy \cite{P,Q,R,S}, Ricci dark energy \cite{T,U,V,W}, agegraphic dark energy \cite{X}, tachyon model \cite{Z,Z1}.

$\bullet$ Dynamical scalar fields: quintessence (or cosmon) \cite{Y.Fujii,Ford,C.Wetterich,B.Ratra,S.M. Carroll,A.Hebecker,A.H,Turner,Caldwell}, ghost condensates \cite{5,6}, phantom \cite{7} and quintom \cite{8}, the model potential is from power-law to exponentials and , to some extent, quintom is an interesting combination of quintessence and phantom.

$\bullet$ Modified gravity: f(R) gravity \cite{ S. Capozziello, S. Capozziello et al, S.M.}, braneworld models \cite{L.,L1,Davli, V. Sahni}, Gauss-Bonnet models \cite{9,10,11,12}, Chern-Simons gravity \cite{13}, Einstein-Aether gravity \cite{14}, cosmological models from scalar-tensor theories of gravity \cite{L. Amendola,J.,T. Chiba,N.Bartolo,F.,V.Sahni and A.A.Starobinsky 2006, P.Ruiz-Lapuente2007}.

In the above, a part of various models on this topic are depicted unsatisfied since there are too many. Hereafter, we plan to focus our attention on the geometric contributions, concretely the so-called Ricci dark energy model based on the holographic principle. Although a complete quantum gravity theory (QGT) has not been developed, we could still explore partly the nature of the dark energy by using the holographic principle which acts as an important result of present QGT (or sting theory) for gravity phenomena. Thus, holographic dark energy model (HDE) constructed in light of the holographic principle can bring us a new perspective about the underlying theory of dark energy. Recently, Gao at al. \cite{W} proposed a new HDE model called Ricci dark energy model (RDE), in which the future event horizon area is replaced by the inverse of the Ricci scalar curvature. They have shown this model does not only avoid the causality problem, phenomenologically viable, but also naturally solves the coincidence problem.

In the present situation, we plan to investigate the astrophysical scales property (wormholes) of the RDE in the universe evolution background and its dependence on the evolution characters of the universe through assuming the dark energy fluid is permeated everywhere. In particular, it is worth noting that, as our previous works \cite{WM1, WM2}, existence of wormholes is always an important problem in physics both at micro and macro scales. There is no doubt that, wormholes together with black holes, pulsars (physical neutron stars) and white dwarfs \cite{zm}, etc., constitute the most attractive, extreme,strange and puzzling astrophysical objects that may provide a new window for physics discovery. Hence, it is necessary to make a brief review about wormholes as follows.

Wormholes could be defined as handles or tunnels in the spacetime topology linking widely separated regions of our universe or of different universes altogether \cite{M.S.Moris and K.S.Thorne1988}. The most fundamental ingredient to form a wormhole is violating the Null Energy Condition (NEC), i.e., $T_{\mu\nu}k^{\mu}k^{\nu}>0$, and consequently all of other energy conditions, where $T_{\mu\nu}$ is the stress-energy tensor and $k^{\mu}$ any future directed null vector.
In general, wormholes in the literature at present can be divided into three classes:

$\bullet$ ordinary wormholes: this class just satisfy the violation of the NEC and is usually not asymptotically flat, singular and consequently non-traversable.

$\bullet$ traversable wormholes: in light of ordinary wormholes, one can obtain the traversable wormholes by an appropriate choice of redshift function or shape function. Subsequently, one can analyze conveniently the traversability conditions of the wormholes and the stabilities.

$\bullet$ thin shell wormholes: one can theoretically construct a geodesically complete traversable wormhole with a shell placed in the junction surface by using the so-called cut-and-paste technique. This class has attracted much attention since the exotic matter required for the existence of spacetime configuration is only located at the shell, and it avoids very naturally the occurrence of any horizon.

Wormholes like other extreme astrophysical objects also have a long theoretical formation history. The earliest remarkable contribution we are aware of is the 1935 introduction of the object now referred to as Einstein-Rosen bridge \cite{Einstein}. Twenty years later, Wheeler first introduced the famous idea of `` spacetime foam '' and coined the term `` wormhole '' \cite{Wheeler}. The actual revival of this field is based on the 1988 paper by Morris and Thorne \cite{M.S.Moris and K.S.Thorne1988}, in which they analyzed for details the construction of the wormhole, energy conditions, time machines, stability problem, and traversabilities of the wormholes. In succession, Visser and Possion introduce the famous `` thin shell wormholes '' by conjecturing that all `` exotic  matter '' is confined to a thin shell between universes \cite{15,16,17,18,19}. After that, there were a lot of papers to investigate the above three classes of wormholes and related properties.

In the past few years, in light of the important discovery that our universe is undergoing a phase of accelerated expansion, an increasing interest to these subjects (wormholes) has arisen significantly in connection with the global cosmology scale discovery. Due to the violation of NEC in both cases (astrophysics wormholes and cosmic dark energy for simple terms), an unexpected and subtle overlap between the two seemingly separated subjects occurs. To be precise, one can usually parameterize the dark energy behaviors by an equation of state of this form $\omega=p/\rho$, where $p$ is the spatially homogeneous pressure and $\rho$ the energy density of the dark energy. In combination with the second Friedmann equation, one can knows that $\omega<-1/3$ is a necessary condition for the cosmic acceleration expansion, $-1<\omega<-1/3$ case is coined to be the quintessence region, $\omega=-1$ is the well-known cosmological constant case (also named phantom divide or cosmic barrier) and  the $\omega<-1$ case corresponds to the phantom region. At the same time, we can easily find that in the phantom range, the NEC is naturally violated. Thus, whatever mysterious dark energy model, if one expects to explore the special wormhole solutions, one must have the phantom case or -like the equation of state properties, from the real cosmic background evolutions with  relevant dark energy models.

In this letter, we plan to investigate the specific geometric RDE wormholes corresponding to a non-ideal equation of state. In particular, traversable wormholes (the second class), whose existence could provide an effective tool for the rapid interstellar travel (another interesting topic for fundamental physics beyond this present work scope), may be of much more interest to causal physics development. So we would like to be aim at studying RDE traversable wormholes since they may contain more constructive physics insights. Especially, it is noteworthy that, the most important result in this letter is that we explore the traversable wormholes constrained by modern astronomical observations, i.e., constraining the equation of state parameters with the cosmic proper evolution stages by using the various astrophysical data-sets (SNe Ia \cite{20}, Baryon Acoustic Oscillations (BAO) \cite{21,22,23,24} and OHD \cite{25,26,27,28,29,30}), which seems to be the first try in the literature. Therefore, one can determine when the exotic matter appears with the evolution of the global universe background in dark energy dominated cosmological models. As a result, one can obtain a substantially clear picture about in which stage of the the evolution of the universe, the wormholes can exist (open) and/or maybe disappear (close) in another different stage. This new connection, to our knowledge, which never be clarified by other authors in the previous literatures about wormholes study, between the wormhole physics and exact cosmology modelings can give us a completely new perspective to investigate the evolution behavior of the wormholes spacetime configurations.

The present paper is organized in the following contexts. In the next section, we make a brief review on the RDE model. In Section. \textrm{III}, we constrain the RDE model by the SNe Ia , BAO and OHD data-sets. In Section. IV, we present a general solution of a traversable wormhole supported by RDE cosmological fluid. In Section. V, we have investigated several specific wormhole geometries and their physical properties and characteristics, including three special choices for redshift function, a specific choice for the shape function, a constant energy density and, finally, isotropic pressure case. In Section. VI, we make a discussion, point out the possible future direction and conclude the present efforts (We adopt the units $8\pi G=c=\hbar=1$ in the following contents).

\section{Review on RDE}
Holographic principle \cite{31, 32} is realized in QGT, which indicates that the entropy of a system increases not with its volume, but with its surface area $L^2$. Cohen et al. \cite{33} proposed an unknown vacuum energy model according to the holographic principle, in which the fine-tuning problem at the cosmological scale as the dark energy and the coincidence problem is also alleviated. But this model has an essential defect, the universe is decelerating and the effective equation of state parameter is zero. Subsequently, Fischler et al. \cite{34,35} proposed the particle horizon could be used as the length scale. Nonetheless, as Hsu \cite{36} and Li \cite{P} pointed out, the equation of state parameter is still greater than $-1/3$, so this model could not explain the expansion mechanism of the universe. For this reason, Li proposed the future event horizon could be used as the characteristic length. This holographic dark energy model could be expressed as follows
\begin{equation}
\rho_H=3c^2M^2_PL^{-2},
\end{equation}
where $\rho_H$ is holographic dark energy density, $c^2$ a dimensionless constant, $M_P$ is the Planck mass and L a box of size containing the total enegy. This model does not only gives an accelerated universe but also is well compatible with the current astronomical observations. Before long, Gao et al. \cite{W} proposed a new HDE model called RDE, in which the future event horizon area is replaced by the inverse of the Ricci scalar curvature. They showed this model does not only avoid the causality problem and is phenomenologically viable, but also naturally solves the coincidence problem. In the following, we consider the spatially flat Friedmann-Robertson-Walker (FRW) universe, and the Ricci scalar curvature is given by
\begin{equation}
R=-6(\dot{H}+2H^2),
\end{equation}
where $H=\dot{s}/s$ is the Hubble parameter, the dot denotes a derivative to the cosmic time $t$. The RDE model states the dark energy density is proportional to the Ricci curvature
\begin{equation}
\rho=\frac{3\alpha}{8\pi}(\dot{H}+2H^2)=-\frac{\alpha}{16\pi}R,
\end{equation}
where $\alpha$ is a constant that can be determined by the current observations. The factor $\frac{3}{8\pi}$ is introduced for simplicity in the following calculations. Combining with the Friedmann equation, Gao et al. \cite{W} obtain the result
\begin{equation}
\rho=3H_0^2[\frac{\alpha}{2-\alpha}\Omega_{m0}e^{-3x}+f_0e^{-(4-\frac{2}{\alpha}x)}],
\end{equation}
where the subscript 0 denotes the present-day value, $\Omega_{m0}\equiv\rho_{m0}/3H_0^2$, $x\equiv\ln{s}$ ($s$ denotes the scale factor) and $f_0$ is an integration constant. Replace Eq.(4) in the energy conservation equation,
\begin{equation}
p=-\rho-\frac{1}{3}\frac{d\rho}{dx},
\end{equation}
one could easily obtain the dark energy pressure
\begin{equation}
p=-3H_0^2(\frac{2}{3\alpha}-\frac{1}{3})f_0e^{-(4-\frac{2}{\alpha}x)}.
\end{equation}
For the convenience of calculations, the Friedmann equation can be rewritten as
\begin{equation}
E^2(z)=\frac{H^2(z)}{H_0^2}=\frac{\alpha}{2-\alpha}\Omega_{m0}(1+z)^3+f_0(1+z)^{2(2-\frac{1}{\alpha})}
\end{equation}
where $E(z)$ denotes the dimensionless Hubble parameter, $z$ the redshift and $z=\frac{1}{s}-1$. In the following, we will constrain the parameters of the RDE model by the SNe Ia, OHD and BAO data-sets.
\section{Astronomical observations Constraints}
\subsection{Type Ia Supernovae}
The observations of SNe Ia provide an forceful tool to probe the expansion history of the universe. As is well known, the absolute magnitudes of all SNe Ia are considered to be the same, since all SNe Ia almost explode at the same mass ($M\approx-19.3\pm0.3$). For this reason, SNe Ia can theoretically be used as the standard candles. In the present letter, we adopt the Union 2.1 data-sets without systematic errors for data fitting, consisting of 580 points covering the range of the redshift $z\in(0.015, 1.4)$. For performing the so-called $\chi^2$ fitting, the theoretical distance modulus is defined as
\begin{equation}
\mu_{th}(z_i)=5\log_{10}D_L(z_i)+\mu_0,
\end{equation}
where $\mu_0=42.39-5\log_{10}h$, $h$ is the dimensionless Hubble parameter today in units of 100 $km^{-1}s^{-1}Mpc$,
\begin{equation}
D_L=(1+z)\int^z_0\frac{dz'}{E(z';\delta)},
\end{equation}
is the Hubble luminosity distance in a spatially flat FRW universe, $\delta$ denotes model parameters. The corresponding $\chi^2_S$ function to be minimized is
\begin{equation}
\chi^2_S=\sum^{580}_{i=1}[\frac{\mu_{obs}(z_i)-\mu_{th}(z_i;\delta)}{\sigma_i}]^2,
\end{equation}
where $\sigma_i$ and $\mu_{obs}(z_i)$ are the corresponding $1\sigma$ error and the observed value of distance modulus for every supernovae. The minimization with respect to $\mu_0$ can be obtained by Taylor-expanding $\chi^2_S$ as \cite{37}
\begin{equation}
\chi^2_S=A-2B\mu_0+C\mu_0^2,
\end{equation}
where
\begin{equation}
A(\delta)=\sum^{580}_{i=1}[\frac{\mu_{obs}(z_i)-\mu_{th}(z_i;\delta;\mu_0=0)}{\sigma_i}]^2,
\end{equation}
\begin{equation}
B(\delta)=\sum^{580}_{i=1}\frac{\mu_{obs}(z_i)-\mu_{th}(z_i;\delta;\mu_0=0)}{\sigma_i^2},
\end{equation}
\begin{equation}
C=\sum^{580}_{i=1}\frac{1}{\sigma_i^2}.
\end{equation}
Therefore, $\chi^2_S$ is minimized when $\mu_0=\frac{B}{C}$ by calculating the transformed $\chi^2_{SN}$ :
\begin{equation}
\chi^2_{SN}=A(\delta)-\frac{[B(\delta)]^2}{C}.
\end{equation}
One can constrain the RDE model by using $\chi^2_{SN}$ which is independent of $\mu_0$ instead of $\chi^2_S$.
\subsection{Observational Hubble Parameter}
In the literature, there are two main methods of independent observational $H(z)$ measurement, which are the `` radial BAO method '' and `` differential age method '' respectively. More details can be found in \cite{38}. The $\chi^2$ for OHD is
\begin{equation}
\chi^2_{O}=\sum^{29}_{i=1}[\frac{H_0E(z_i)-H_{obs}(z_i)}{\sigma_i}]^2.
\end{equation}
Using the same trick in the above, the minimization with respect to $H_0$ can be made trivially by Taylor-expanding $\chi_{OHD}^2$ as
\begin{equation}
\chi^2_{O}(\delta)=AH_0^2-2BH_0+C,
\end{equation}
where
\begin{equation}
A=\sum^{29}_{i=1}\frac{E^2(z_i)}{\sigma_i^2},
\end{equation}
\begin{equation}
B=\sum^{29}_{i=1}\frac{E(z_i)H_{obs}(z_i)}{\sigma_i^2},
\end{equation}
\begin{equation}
C=\sum^{29}_{i=1}\frac{H^2_{obs}(z_i)}{\sigma_i^2}.
\end{equation}
Therefore, $\chi^2_O$ is minimized when $H_0=\frac{B}{A}$ by calculating the following transformed $\chi^2_{OHD}$ :
\begin{equation}
\chi^2_{OHD}=-\frac{B^2}{A}+C.
\end{equation}
One can constrain the RDE model by using $\chi^2_{OHD}$ which is independent of $H_0$ instead of $\chi^2_O$.
\subsection{Baryon Acoustic Oscillations}
\begin{figure}
\centering
\includegraphics[scale=0.5]{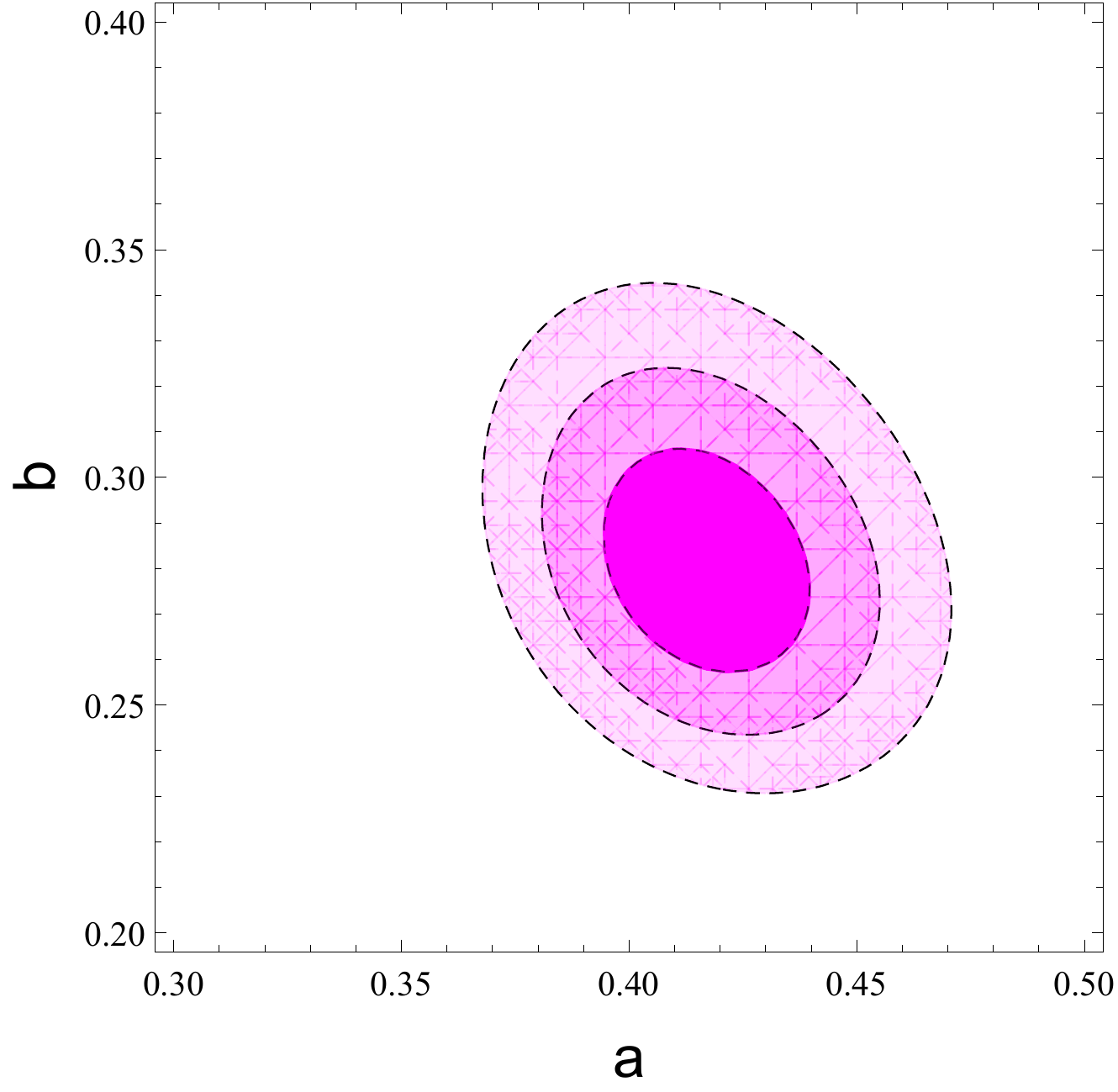}
\caption {1$\sigma$, 2$\sigma$ and 3$\sigma$ confidence ranges for parameter pair ($a$, $b$) of RDE model, constrained by SNe Ia and BAO data-sets (Here we fix the parameter $c$ at the best fitting value $c=0.644957$). For simplicity, we denote the model parameters $\alpha=a$, $\Omega_{m0}=b$ and $f_0=c$, respectively.}\label{fig1}
\end{figure}

\begin{figure}
\centering
\includegraphics[scale=0.5]{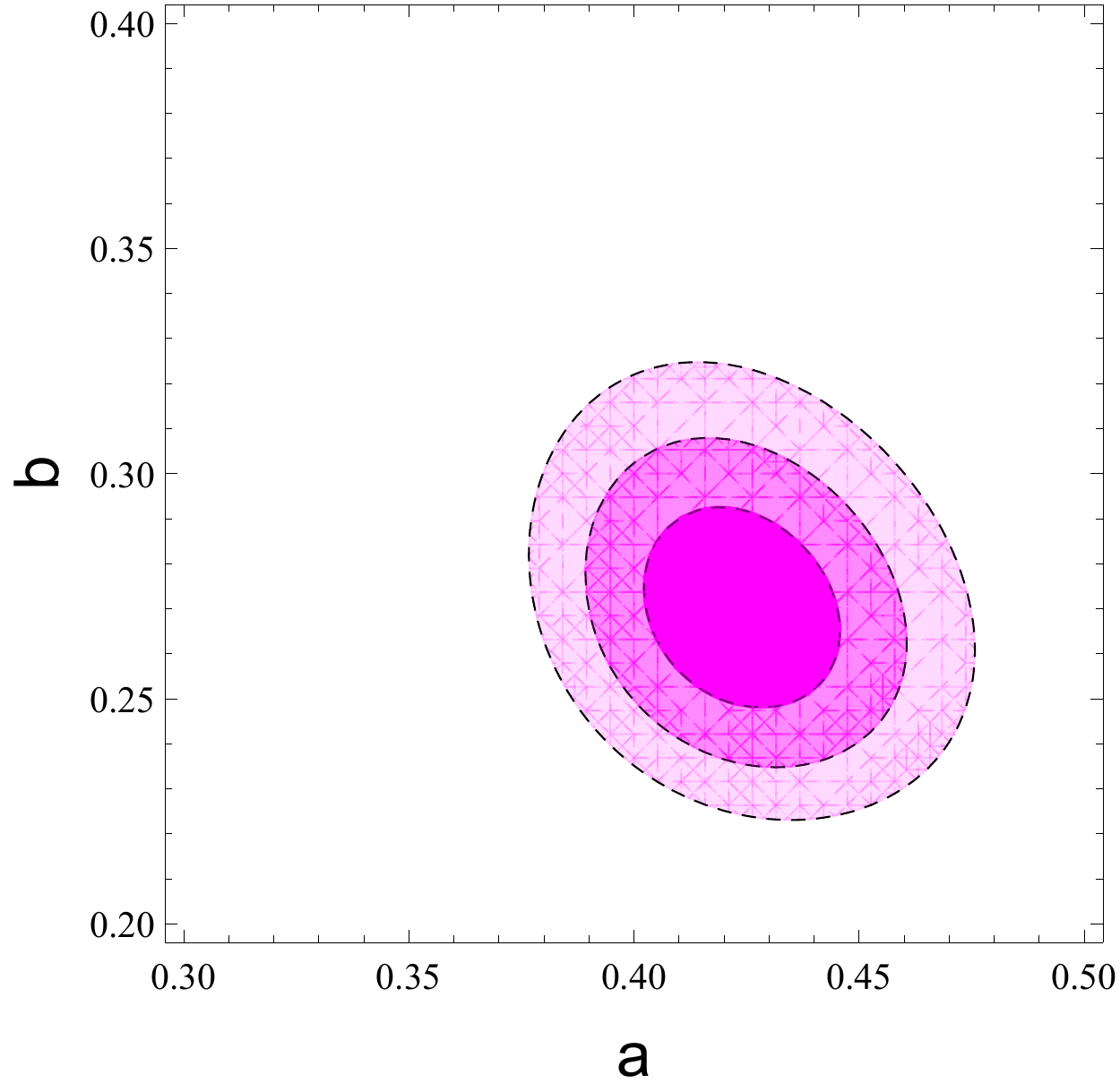}
\caption{1$\sigma$, 2$\sigma$ and 3$\sigma$ confidence ranges for parameter pair ($a$, $b$) of RDE model, constrained by SNe Ia, OHD and BAO data-sets (Here we fix the parameter $c$ at the best fitting value $c=0.650594$). For simplicity, we denote the model parameters $\alpha=a$, $\Omega_{m0}=b$ and $f_0=c$, respectively. \label{fig2}}
\end{figure}

\begin{figure}
\centering
\includegraphics[scale=0.5]{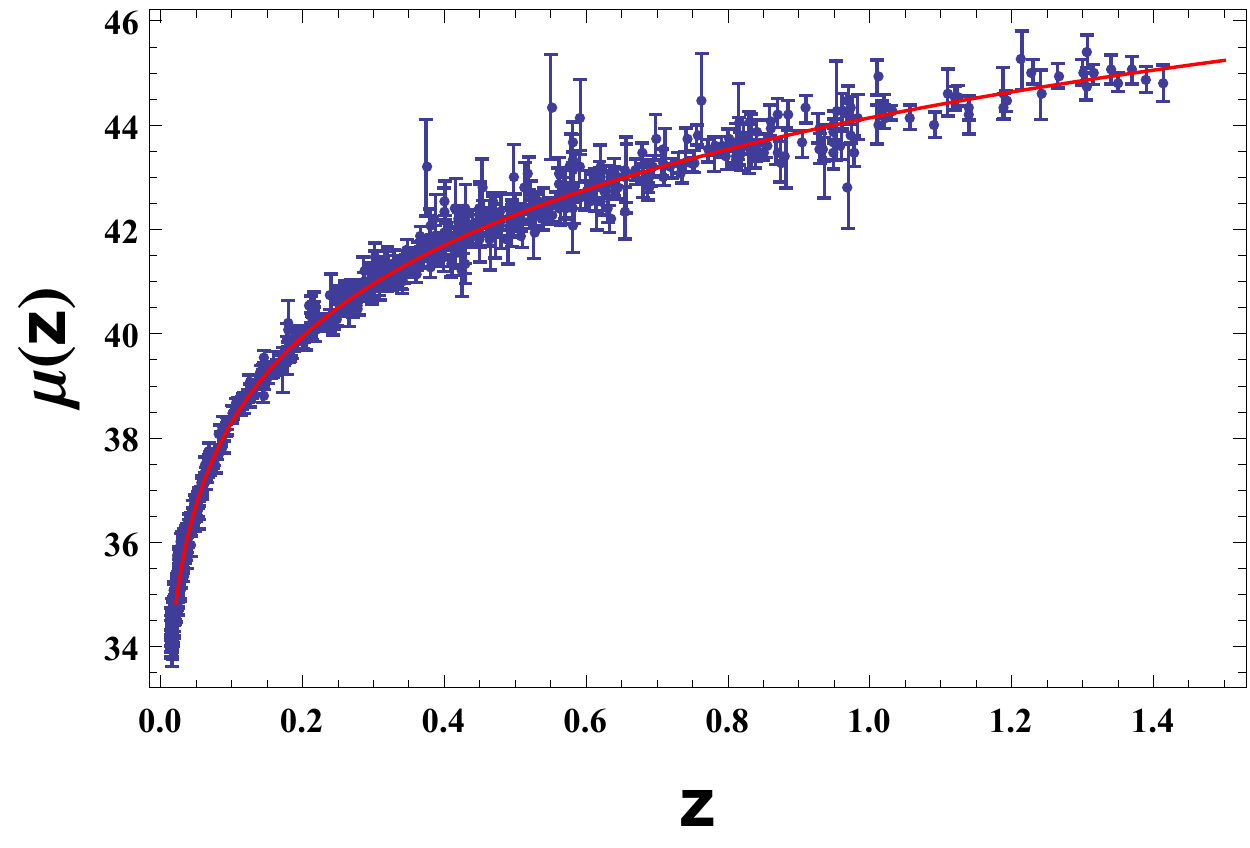}
\caption {The relation between distance modulus and redshift. The solid line represents the theoretical curve calculated from the model concerned. The dots with errors bar correspond to the 580 data points from the supernovae observations. One can find that the theoretical curve is well consistent with the observations.} \label{fig3}
\end{figure}

\begin{figure}
\centering
\includegraphics[scale=0.5]{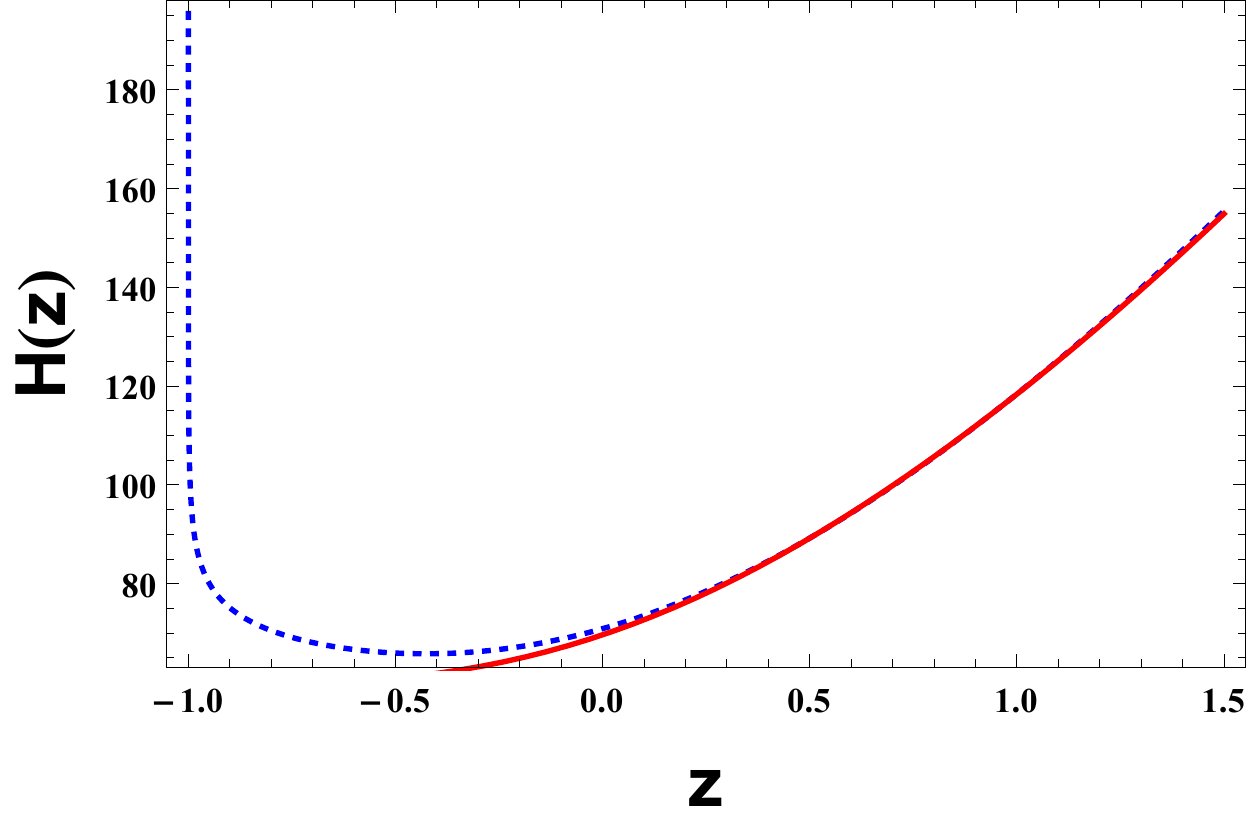}
\caption{The relation between the Hubble Parameter and redshift. The solid (red) line and the dotted (blue) line correspond to the $\Lambda$CDM model and the RDE model, respectively. One can easily find that the RDE model is well consistent with the $\Lambda$CDM model in the past. Nonetheless, when the redshift $z\rightarrow0$ gradually, one may discover that the discrepancy occurs: the Hubble parameter of the RDE model will be a little higher than the standard cosmological model. In the remote future, the expansion velocity of the universe in RDE model will diverge, i.e., the universe tends to be a super rip.}\label{fig4}
\end{figure}

In addition to the SNe Ia and OHD data-sets, another constraint is from BAO traced by the Sloan Digital Sky Survey (SDSS). We use the distance parameter $\mathcal{A}$ to measure the BAO peak in the distribution of SDSS luminous red galaxies, and the distance parameter $\mathcal{A}$ can be defined as
\begin{equation}
\mathcal{A}=\sqrt{\Omega_{m0}}E(z_a)^{-\frac{1}{3}}[\frac{1}{z_a}\int^{z_a}_0\frac{dz'}{E(z')}]^{\frac{2}{3}},
\end{equation}
where $z_a=0.35$. The $\chi^2$ for BAO data-sets is
\begin{equation}
\chi^2_{BAO}=\sum^6_{i=1}[\frac{\mathcal{A}_{obs}(z_i)-\mathcal{A}_{th}(z_i;\delta)}{\sigma_{\mathcal{A}}}]^2.
\end{equation}
In the following, for simplicity, we denote the model parameters $\alpha=a$, $\Omega_{m0}=b$ and $f_0=c$, respectively. At first, we compute the joint constraints from SNe Ia and BAO data-sets. The $ \chi^2$ can be defined as
\begin{equation}
\chi^2_{1}=\chi^2_{SN}+\chi^2_{BAO}.
\end{equation}
In the second place, we calculate the combined constraints from SNe Ia, OHD and BAO data-sets. The $\chi^2$ can be defined as
\begin{equation}
\chi^2_{2}=\chi^2_{SN}+\chi^2_{BAO}+\chi^2_{OHD}.
\end{equation}
The likelihoods of the parameters ($a$, $b$) in the two different constraints ($\chi^2_{1}$ and $\chi^2_{2}$) are depicted in Fig. \ref{fig1} and Fig. \ref{fig2}, respectively. The best fitting values of the parameters and the values of the reduced $\chi^2_{1}$ and $\chi^2_{2}$ are listed in Table. \ref{tab1}. At the same time, it is very constructive to show the relation between the distance modulus and redshift (Fig. \ref{fig3}). As a result, one can naturally get the evolution behavior of the universe when taking the parameters as the best fitting values (Fig. \ref{fig4}). In addition, the effective equation of state parameter $\omega_X$ with respect to the redshift $z$ from data fitting (see Table. \ref{tab1}) are depicted in Fig.5.
\begin{figure}
\centering
\includegraphics[scale=0.5]{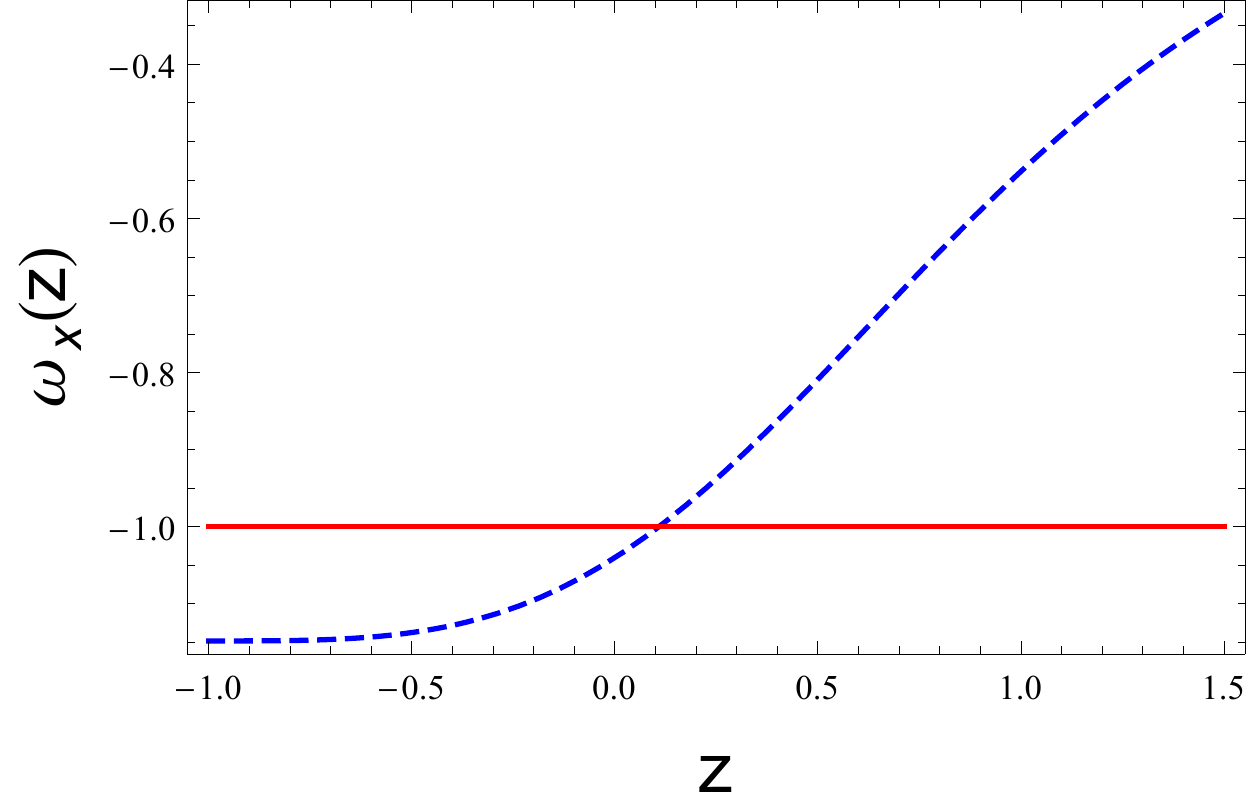}
\caption{The relation between the effective equation of state parameter $\omega_X(z)$ and the redshift $z$: the solid (red) line and the dotted (blue) line correspond to the $\Lambda$CDM model and the RDE model, respectively. One can apparently discover that the RDE model represents a quintom-like model crossing the phantom divide $\omega_X(z)=-1$, which mainly include two types of exotic matter: the phantom-like and the quintessence-like. It is noteworthy that when $\omega_X(z)<-1$, the wormholes will appear in the universe (open) and when $z\rightarrow-1$, the wormholes will disappear (close) since the universe tends to a super rip.}\label{fig5}
\end{figure}
\begin{table}[h!]
\caption{The best fitting values of the parameters for the RDE model by using two different kinds of joint constraints: 580 SNe Ia gold samples, 7 BAO data points and 29 OHD data points.}
\label{tab1}
\begin{tabular}{ccccc}
\hline
SNe Ia+BAO& &SNe Ia+OHD+BAO&\\
\hline
$\chi^2_{min}/d.o.f.$ & $566.071/(587)$ &$564.098/(616)$\\
$a$  & $0.4164098$ & $0.423469$ \\
$b$  & $0.280558$& $0.268549$ \\
$c$  & $0.644957$& $0.650594$ \\
\hline
\end{tabular}
\end{table}

In Fig. \ref{fig3}, one can easily get the conclusion that the theoretical curve of the distance modulus with respect to redshift is well behaved by a comparison with the 580 SNe Ia samples. In Fig. \ref{fig4}, one can find that the cosmological background evolution of the RDE model is consistent with the $\Lambda$CDM model in the past. However, when $z$ approaches 0 corresponding the present universe, the discrepancy occurs, the Hubble parameter of the RDE model will be a little higher than the standard cosmological model. In the future, the expansion velocity of the universe in RDE model will diverge, namely, the universe tends to be a super rip. In Fig. \ref{fig5}, one can not only find that the evolution behavior of RDE model, but also apparently discover that the change of the type of the cosmic matter (phantom-like or qiuntessence-like) with the evolution of the universe by comparing with the $\Lambda$CDM model. One can easily get that the RDE model corresponds to a quintom-like matter (Virtually, still phantom-like or qiuntessence-like) . It is worth noting that, this is the essential starting point of our work. Since we can describe quantitatively the evolution of the type of the cosmic matter by astrophysical observations, we can explore the related physics for a concrete type of cosmic matter.

In particular, we are of much interest in the attractive and elegant objects, wormholes. Hence, in this letter, we introduce astrophysical observations into the field of wormhole physics which seems to be the first try in the literature. Through the astronomical observations, we make a constraint on the parameters of a cosmological model, explore the type of cosmic matter, limit the number of available models for wormhole research, reduce the number of the wormholes corresponding to different parameters for a concrete cosmological model and provide a more clear picture for wormhole research from the point of view of observation cosmology. For an illustration, in the following, we investigate the traversable wormholes in the RDE model.
\section{RDE traversable wormholes}
\subsection{Basic Equations}
In the present letter, we consider the spacetime geometry representing a static and spherically symmetric wormhole
\begin{equation}
ds^2=-e^{2\Phi(r)}dt^2+\frac{dr^2}{1-\frac{b(r)}{r}}+r^2(d\theta^2+\sin^2\theta d\phi^2),
\end{equation}
where $b(r)$ and $\Phi(r)$ are arbitrary functions of the radial coordinate $r$, denoted as the shape function and redshift function, respectively \cite{M.S.Moris and K.S.Thorne1988}. The radial coordinate $r$ runs in the range $r_0\leq r<\infty$ where $r_0$ corresponds to the radius of the wormhole throat, $0\leq\theta\leq\pi$ and $0\leq\phi\leq2\pi$ are the angular coordinates. One can also consider a cutoff of the stress energy tensor at a junction radius $a$.

As mentioned above, the most fundamental requirement to form a wormhole is violating the NEC. Besides, there are also two fundamental requirements to form a traversable wormhole. The first fundamental ingredient is satisfying the so-called flaring out condition that can be expressed as follows:
\begin{equation}
b(r_0)=r_0,
\end{equation}
\begin{equation}
b'(r_0)<1,
\end{equation}
\begin{equation}
b(r)<r,r>r_0.
\end{equation}
Another fundamental ingredient of a traversable wormhole is that $\Phi(r)$ must be finite everywhere, in order to avoid an horizon, which can be identified the surfaces with $e^{2\Phi(r)}\rightarrow0$. For a wormhole geometry to be asymptotically flat, one also demands that $b/r\rightarrow0$ and $\Phi\rightarrow0$ as $r\rightarrow\infty$. In the next subsection, we will show the cutoff of the stress energy tensor.

Using the Einstein Field Equations, $G_{\mu\nu}=T_{\mu\nu}$, one can obtain the following relationships:
\begin{equation}
b'=r^2\rho,
\end{equation}
\begin{equation}
\Phi'=\frac{b+r^3p_r}{2r^2(1-b/r)},
\end{equation}
where the prime denotes a derivative with respect to the radial coordinate $r$, $\rho(r)$ is the matter energy density and $p_r(r)$ is the radial pressure of the cosmic fluid. One could also derive from the conservation law of the stress-energy tensor $T^{\mu\nu}_{\hspace{3mm};\nu} = 0$ with $\mu=r$ that
\begin{equation}
p'_r=\frac{2}{r}(p_t-p_r)-(\rho+p_r)\Phi',
\end{equation}
where $p_t(r)$ represents the lateral pressure measured in the orthogonal direction to the radial direction. Eq.(7) can also be interpreted as the relativistic Euler equation or the hydrostatic equation for equilibrium for the material threading the wormhole.

Combining Eq. (4) and Eq. (6), one can easily derive that the equation of state of RDE model,
\begin{equation}
p=-(\frac{2}{3\alpha}-\frac{1}{3})(\rho-\frac{3\alpha H_0^2\Omega_{m0}e^{-3x}}{2-\alpha}).
\end{equation}
For the conveniences of calculations, we denote $\eta=H_0^2\Omega_{m0}e^{-3x}$ in the following. Moreover, we will consider the pressure in the RDE equation of state is the radial pressure, therefore, Eq. (33) can be rewritten as
\begin{equation}
p_r=-(\frac{2}{3\alpha}-\frac{1}{3})(\rho-\frac{3\alpha\eta}{2-\alpha}).
\end{equation}
Using the Eq. (30) and Eq. (31), we can obtain the following relation:
\begin{equation}
\Phi'(r)=\frac{b-r^3(\frac{2}{3\alpha}-\frac{1}{3})(\frac{b'}{r^2}-\frac{3\alpha\eta}{2-\alpha})}{2r^2(1-\frac{b}{r})}.
\end{equation}
One can denote the solutions of Eq. (26) satisfying Eq. (35) as `` RDE wormholes ''. Furthermore, if the solutions also satisfy the traversability condition, we denote it as `` RDE traversable wormholes ''.

Subsequently, we apply the condition Eq. (28) into the RDE equation of state evaluated at the throat. We verify that energy density at $r_0$ is $\rho(r_0)=\frac{3\alpha(1+\eta r_0^2)}{r_0^2(2-\alpha)}$. Then, using Eq. (30) and the condition Eq. (28), we obtain the following relation:
\begin{equation}
\eta-\frac{2(1-2\alpha)(1+\eta r_0^2)}{r_0^2(2-\alpha)}<0.
\end{equation}
One can also get the same conclusion from the violation of the NEC at the wormhole throat, namely, $p_r(r_0)+\rho(r_0)<0$.
\subsection{Theoretical construction of asymptotically flat spacetime}
As mentioned above, one can construct an asymptotically flat spacetime, in which $b/r\rightarrow0$ and $\Phi\rightarrow0$ as $r\rightarrow\infty$. In general, it is difficult to obtain a flat spacetime directly. So one may construct theoretically solutions by matching the interior geometry into an exterior vacuum geometry. If the surface stresses at the matching radius is zero, we call it boundary surface. To the opposite, if the surface stresses is present, we denote it as a thin shell \cite{39,40}.

For simplicity, we just consider a simple exterior spacetime solution, namely, the Reissner-Norsdtr\"{o}m spacetime
\begin{equation}
ds^2=-(1-\frac{2M}{r}+\frac{Q^2}{r^2})dt^2+\frac{dr^2}{1-\frac{2M}{r}+\frac{Q^2}{r^2}}+r^2(d\theta^2+\sin^2\theta d\phi^2).
\end{equation}
where Q is the charge. For $\left|Q\right|<M$ this geometry has an inner and outer (event) horizon given by
\begin{equation}
r_{\pm}=M\pm\sqrt{M^2-Q^2},
\end{equation}
if $\left|Q\right|=M$ the two horizons merge into one, and when $\left|Q\right|>M$ there are no horizons and the metric represents a naked singularity. Particularly, when $\left|Q\right|\leq M$ the radius of the wormhole throat $r_0$ should be taken greater than $r_h=r_+$ (Here $r_h$ represents the event horizon), i.e., $a>r_+$ (here $a$ is the junction radius), in order that no horizons are present in the whole spacetime. If $\left|Q\right|>M$ the condition $r_0>0$ naturally ensures that the naked singularity is removed.

Using the Darmois-Israel formalism \cite{41,42}, one can find the surface stresses of a dynamical thin shell surrounding the wormhole are given by
\begin{equation}
\sigma=-\frac{1}{4\pi a}(\sqrt{1-\frac{2M}{a}+\frac{Q^2}{a^2}+\dot{a}^2}-\sqrt{1-\frac{b(a)}{a}+\dot{a}^2}),
\end{equation}
\begin{equation}
P=\frac{1}{8\pi a}[\frac{1-\frac{M}{a}+\dot{a}^2+a\ddot{a}}{\sqrt{1-\frac{2M}{a}+\frac{Q^2}{a^2}+\dot{a}^2}}-\frac{(1+a\Phi')(1-\frac{b}{a}+\dot{a}^2)+a\ddot{a}-\frac{\dot{a}^2(b-b'a)}{2(a-b)}}{\sqrt{1-\frac{b(a)}{a}+\dot{a}^2}}],
\end{equation}
where the overdot denotes the derivative with respect to the proper time $\tau$, $\sigma$ and $P$, respectively, denote the surface energy density and the lateral surface pressure. One may expect to obtain the static thin shell formalism of the above metric, so by taking into account $\dot{a}=\ddot{a}=0$, we can get

Through doing the same as the above case, we obtain the surface stresses of the static thin shell as follows:
\begin{equation}
\sigma=-\frac{1}{4\pi a}(\sqrt{1-\frac{2M}{a}+\frac{Q^2}{a^2}}-\sqrt{1-\frac{b(a)}{a}}),
\end{equation}
\begin{equation}
P=\frac{1}{8\pi a}[\frac{1-\frac{M}{a}}{\sqrt{1-\frac{2M}{a}+\frac{Q^2}{a^2}}}-\frac{(1+a\Phi')(1-\frac{b}{a})}{\sqrt{1-\frac{b(a)}{a}}}],
\end{equation}
thus, we have obtained the surface energy density and the tangential pressure of static thin shell in the simplest case with charge.
\section{Exact solutions}
\subsection{Constant redshift function}
For a constant redshift function, namely, $\Phi'=0$, one can get the following shape function:
\begin{equation}
b(r)=r_0(\frac{r}{r_0})^{\frac{3\alpha}{2-\alpha}}+\frac{\alpha\eta}{2(1-\alpha)}[r^3-r_0^3(\frac{r}{r_0})^{\frac{3\alpha}{2-\alpha}}].
\end{equation}
It is easy to be seen that $b(r)<r$ which satisfies the inequality (29). Evaluating at the throat, it follows that
\begin{equation}
b'(r_0)=\frac{3\alpha}{2-\alpha}(1+\eta r_0^2).
\end{equation}
Here we adopt the best fitting values of the parameters in Table. I, and the inequality (28) is well satisfied. For instance, $b'(r_0)\approx0.805824<1$ when we use the best fitting values of SNe Ia, OHD and BAO data-sets. Interestingly, this solution is not only traversable but also asymptotically flat since $\Phi$ is finite and $b/r\rightarrow0$ when $r\rightarrow\infty$. Therefore, the dimensions of the wormhole can be substantially large.

In wormhole physics, the most fascinating thing may be to analyze the traversabilities including traversal velocity and traversal time for a human being to journey through the wormhole. In general, there are three constraint conditions. The first one is that the acceleration felt by the traversers should not exceed 1 Earth's gravity $g_\oplus$ (see \cite{M.S.Moris and K.S.Thorne1988} for more details)
\begin{equation}
\left|(1-\frac{b}{r})^{\frac{1}{2}}(\gamma e^\Phi)'e^{-\Phi}\right|\leq g_\oplus.
\end{equation}
where $\gamma=(1-v^2)^{-1/2}$. The second one is that the tidal acceleration also should not exceed 1 Earth's gravitational acceleration:
\begin{equation}
\left|\lambda^1\right|\left|(1-\frac{b}{r})[\Phi''+(\Phi')^2-\frac{b'r-b}{2r(r-b)}\Phi']\right|\leq g_\oplus,
\end{equation}
\begin{equation}
\left|\lambda^2\right|\left|\frac{\gamma^2}{2r^2}[v^2(b'-\frac{b}{r})+2(r-b)\Phi']\right|\leq g_\oplus,
\end{equation}
where $v$ is the traveler's velocity and $\left|\delta^i\right|$ is the distance between two arbitrary parts of the traveler's body (the size of the traveler). According to \cite{M.S.Moris and K.S.Thorne1988}, we should assume $\left|\lambda^i\right|\approx2 m$ along any spatial direction in the traveler's reference frame. The last condition that the traversal time measured by the traveler and for the observers who remain at rest at space stations are, respectively, given by
\begin{equation}
\Delta\tau=\int^{+l_2}_{-l_1}\frac{dl}{v\gamma},
\end{equation}
\begin{equation}
\Delta t=\int^{+l_2}_{-l_1}\frac{dl}{ve^\Phi},
\end{equation}
where $dl=(1-\frac{b}{r})^{-1/2}dr$ is the proper radial distance, and we consider that the space stations are located at a radius $r=a$, at $-l_1$ and $l_2$, respectively.

It is obvious that inequalities (45) and (46) is well satisfied in the case of a constant redshift function. Substituting Eqs. (43-44) into inequality (47) evaluated at the throat, neglect the substantial small term that contains the redshift $z$, considering a constant non-relativistic traversal velocity, namely, $\gamma\approx1$, we obtain the new conclusion for the velocity:
\begin{equation}
v\leq r_0\sqrt{\frac{(2-\alpha)g_\oplus}{(1-2\alpha)\left|\lambda^2\right|}}.
\end{equation}
In the following, through considering the equality case, assuming that the throat radius is given by $r_0\approx100$ m and taking into account the best fitting value $\alpha=0.423469$, we get the traversal velocity $v\approx710.42$ m/s. Furthermore, if one take into consideration the matching radius is provided by $d=10000m$, then, one can obtain $\Delta\tau\approx\Delta t\approx28.15$ s from the traversal times $\Delta\tau\approx\Delta t\approx2a/v$ (one can compare this case with that in \cite{39} in order to get more useful information).
\subsection{$\Phi(r)=\ln(\frac{r_0}{r})$}
One can also make another choice of the redshift function that seems to be a little more complex than the first case, i.e., $\Phi(r)=\ln(\frac{r_0}{r})$. Substituting it into Eq. (35), one can get
\begin{equation}
-\frac{1}{r}=\frac{b-r^3(\frac{2}{3\alpha}-\frac{1}{3})(\frac{b'}{r^2}-\frac{3\alpha\eta}{2-\alpha})}{2r^2(1-\frac{b}{r})}.
\end{equation}
Solving this equation, it follows that
\begin{equation}
b(r)=r_0(\frac{r}{r_0})^{\frac{3\alpha}{\alpha-2}}(\frac{1-2\alpha}{1+\alpha}-\frac{\alpha\eta r_0^2}{2})+\frac{3\alpha}{1+\alpha}r+\frac{\alpha\eta r^3}{2}
\end{equation}
Furthermore, by differentiating both sides with respect to $r$, one can have
\begin{equation}
b'(r)=\frac{3\alpha}{\alpha-2}(\frac{r}{r_0})^{\frac{2(2\alpha-1)}{2-\alpha}}(\frac{1-2\alpha}{1+\alpha}-\frac{\alpha\eta r_0^2}{2})+\frac{3\alpha}{1+\alpha}+\frac{3\alpha\eta^2r^2}{2}.
\end{equation}
It is easy to be checked that the shape function satisfies the flaring out conditions and $b'(r_0)$ is still the same with the case of constant redshift function (Eq.(44)). Unfortunately, this solution is not asymptotically flat. However, one can glue it to an exterior vacuum spacetime at a matching radius $a$. Moreover, this solution is a traversable wormhole since the corresponding redshift function is finite in the range $r_0\leq r\leq a$.

Before ten years, Visser et al. discovered that one could theoretically construct traversable wormholes with infinitesimal amounts of average null energy conditions (ANEC) violating matter \cite{43,44}. To be precise, taking into consideration the notion of `` volume integral quantifier '', one could quantify the total amounts of exotic matter by computing the definite integrals $\int T_{\mu\nu}U^\mu U^\nu dV$ and $\int T_{\mu\nu}k^\mu k^\nu dV$, and the amount of exotic mater is defined as how negative the values of these integrals become. It is of much interest for us to apply this method into the RDE model, in order to study whether it is the same case. Then, using this effective method, given by $I_V=\int[p_r(r)+\rho]dV$, with a cutoff of the stress energy tensor at $a$, one can obtain
\begin{equation}
I_V=[(r-b)\ln(\frac{e^{2\Phi}}{1-\frac{b}{r}})]^a_{r_0}-\int^a_{r_0}(1-b')[\ln(\frac{e^{2\Phi}}{1-\frac{b}{r}})]dr
=\int^a_{r_0}(r-b)[\ln(\frac{e^{2\Phi}}{1-\frac{b}{r}})]'dr.
\end{equation}
where we have considered the asymptotical flat case, so the first boundary term vanishes. Then, the above integral can be calculated out, by using the mentioned-above redshift function and shape function, as follows(for simplicity, neglect the term containing redshift $z$):
\begin{equation}
I_V=\frac{2(2\alpha-1)[3a\alpha-(\alpha+1)r_0-(\frac{a}{r_0})^{\frac{3\alpha}{\alpha-2}}(2\alpha-1)]}{3\alpha(1+\alpha)}.
\end{equation}
It is worth noting that if we adopt the best fitting parameter $\alpha=0.423469$ from the astrophysical observations, the integral can be expressed as
\begin{equation}
I_V=-0.16928[1.27041a-1.42347r_0+0.153062r_0(\frac{r_0}{a})^{0.805824}].
\end{equation}
It is easy to be seen that when taking the limit $a\rightarrow r_0$ the integral will be zero, namely, $I_V\rightarrow0$. This indicates that we can theoretically construct a traversable wormhole with infinitesimal amounts of ANEC violating RDE matter. In addition, one can find that this method may provide more information for us about the `` total amount '' of ANEC violating matter in the whole spacetime (see \cite{43} for more details).
\subsection{$\Phi(r)=\ln(\frac{r}{r_0})$}
As a comparison with the case $\Phi(r)=\ln(\frac{r_0}{r})$, we consider $\Phi(r)=\ln(\frac{r}{r_0})$ here. Similarly, one obtains
\begin{equation}
b(r)=r_0(\frac{r}{r_0})^{\frac{9\alpha}{2-\alpha}}[1-\frac{3\alpha}{5\alpha-1}-\frac{\alpha\eta r_0^3}{2(1-2\alpha)}]+\frac{3\alpha}{5\alpha-1}r+\frac{\alpha\eta r^3}{2(1-2\alpha)}
\end{equation}
and
\begin{equation}
b'(r)=\frac{9\alpha}{2-\alpha}(\frac{r}{r_0})^{\frac{2(5\alpha-1)}{2-\alpha}}[1-\frac{3\alpha}{5\alpha-1}-\frac{\alpha\eta r_0^3}{2(1-2\alpha)}]+\frac{3\alpha}{5\alpha-1}+\frac{3\alpha\eta r^2}{2(1-2\alpha)}.
\end{equation}
Evaluating at the throat, one get the same expression as Eq. (44) again:
\begin{equation}
b'(r_0)=\frac{3\alpha}{2-\alpha}(1+\eta r_0^2).
\end{equation}
Obviously, this solution reflects a non-asymptotically flat wormhole. Adopting the same step as mentioned-above, we can also construct a traversable wormhole in the finite range. It is more noteworthy that, we think, the three choices of the redshift function can give us a mathematical paradigm as Eq. (44). This means, at the throat $r=r_0$, the violation of the NEC can provide the same result for us ($p_r(r_0)+\rho(r_0)<0$). Furthermore, these three wormholes have a high degeneracy at the same throat. More physically, when three different travelers cross the throats of three different wormholes at the same moment, respectively, they may see the same bending of light and feel the same radial pressure, gravitation acceleration, etc. In addition, if we take the parameter $\alpha=0.423469$ once again, then, $b'(r_0)\approx0.805824<1$. Therefore, the introduction of the astrophysical observations will provide more useful information to investigate the behavior of the objects.
\subsection{$b(r)=r_0(\frac{r}{r_0})^\epsilon$}
Take into account the case $b(r)=r_0(\frac{r}{r_0})^\epsilon$, one can obtain the redshift function as follows from Eq. (35):
\begin{equation}
\Phi'(r)=\frac{r_0(\frac{r}{r_0})^\epsilon-r^3(\frac{2}{3\alpha}-\frac{1}{3})[\frac{\epsilon}{r^2}(\frac{r}{r_0})^{\epsilon-1}-\frac{3\alpha\eta}{2-\alpha}]}{2r^2[1-(\frac{r_0}{r})^{1-\epsilon}]}
\end{equation}
Unfortunately, this equation can not be solved analytically. Thus, one can solve it numerically for every given parameter $\epsilon$. Moreover, by using inequality (28), one can discover that $\epsilon<1$. In the following, we only consider the particular case $\epsilon=\frac{1}{2}$. So the redshift function becomes
\begin{equation}
\Phi(r)=\frac{r\eta[6r_0+3r+4\sqrt{\frac{r_0}{r}}(3r_0+r)]-(7\alpha-2)[2arctanh{\sqrt{\frac{\alpha r_0}{(1+\alpha)r}}+\ln(1-\frac{r_0}{r})]}+6r_0^2\eta[\ln(\frac{r}{r_0})-Beta[\frac{r}{r_0},\frac{1}{2},1]]}{12\alpha},
\end{equation}
where $Beta[\frac{r}{r_0},\frac{1}{2},1]$ is the incomplete Beta function, which equals 2 evaluated at the throat $r=r_0$. Through some simple calculations, one can find that $\Phi(r)$ is finite everywhere. Nonetheless, this solution is not asymptotically flat since $\Phi(r)\rightarrow\infty$ when $r\rightarrow\infty$. Similarly, one can construct a traversable wormhole by matching the interior geometry into an exterior vacuum geometry.
\subsection{Constant dark energy density}
\begin{figure}
\centering
\includegraphics[scale=0.5]{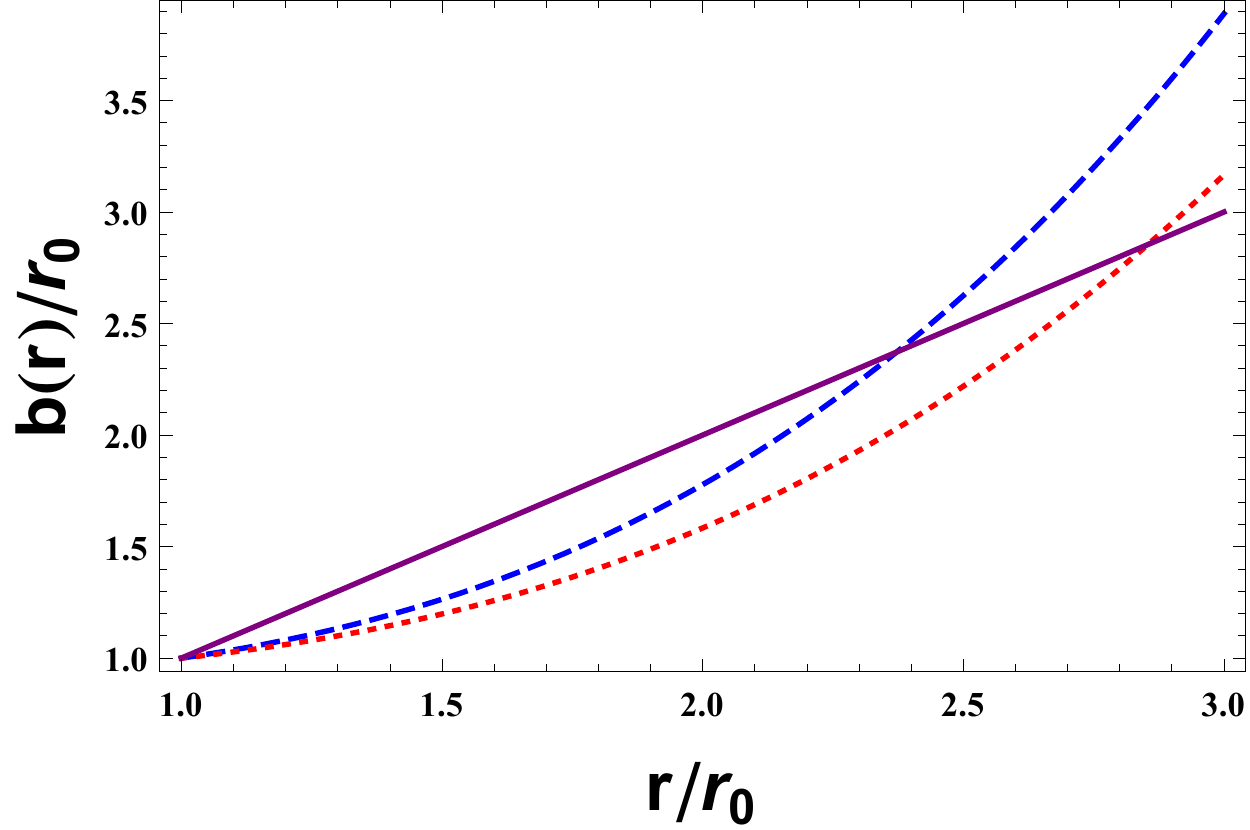}
\caption{To form a wormhole geometry, according to the restriction $b(r)<r$, one can find that the range below the solid (purple) line represents  wormhole solutions. The solid line corresponds to the line of $b(r)=r$, the dashed (blue) line corresponds to the line of $\beta=1/3$ and the dotted (red) line corresponds to the line of $\beta=1/4$. In addition, one may discover that the dimensions of wormholes rely on the values of $\beta$, namely,  the wormhole dimensions decrease when the values of $\beta$ increase.}\label{fig6}
\end{figure}

Taking into account a constant dark energy density as in \cite{45}, namely, $\rho=\rho_0$, from Eq.(30) one can obtain
\begin{equation}
b(r)=\frac{\rho_0}{3}(r^3-r_0^3)+r_0.
\end{equation}
By a new definition $A=\frac{\rho_0}{3}$, Eq. (28) can be expressed as: $3Ar_0^2<1$. Take into consideration $A=\frac{\beta}{3r_0^2}$, with $0<\beta<1$, in order that Eq. (62) can be rewritten as
\begin{equation}
b(r)=r_0\{\frac{\beta}{3}[(\frac{r}{r_0})^3-1]+1\}.
\end{equation}
To form a wormhole geometry, Eq.(29) must be satisfied. Through some calculations, one can find that $b(r)=r$ has two positive roots: $r_1=r_0$ and $r_2=r_0\frac{\sqrt{\frac{12}{\beta}-3}-1}{2}$, and $r$ lies in the finite range
\begin{equation}
r_0<r<r_0\frac{\sqrt{\frac{12}{\beta}-3}-1}{2}.
\end{equation}
To be more clear, this constrain condition is shown graphically in Fig. \ref{fig6}. One can get the conclusion that the dimensions of wormholes decreases when the values of $\beta$ increase.
Substituting Eq. (63) into Eq. (35), one can find that the redshift function can be expressed as
\begin{equation}
\begin{aligned}
\Phi(r)=&C_1+\frac{1}{2\sqrt{3}(\beta-1)\sqrt{\beta(1-\beta)}}arctanh[\frac{\beta(2r+r_0)}{r_0\sqrt{3\beta(4-\beta)}}][3(\alpha\beta+\eta r_0^2)+(\alpha-2)r_0^2\rho_0]-\frac{1}{2}\ln(r)\\
&+\frac{\ln[(\beta-3)r_0^2+\beta r_0r+\beta r^2]\{r_0^2(2\beta-3)[3\eta+(\alpha-2)\rho_0]-3\alpha\beta\}}{12\alpha\beta(1-\beta)}+\frac{\ln(r-r_0)[3(\alpha+r_0^2\eta)+(\alpha-2)r_0^2\rho_0]}{6(1-\beta)},
\end{aligned}
\end{equation}
where $C_1$ is an integration constant. One may find when $r=r_0$ there is an event horizon, so the solution is a non-traversable wormhole. Nonetheless, assuming the condition $\rho_0=\frac{3(\alpha+\eta r_0^2)}{(2-\alpha)r_0^2}$, Eq. (65) will reduce to
\begin{equation}
\begin{aligned}
\Phi(r)=&C_1+\frac{\sqrt{3}}{2(\beta-1)\sqrt{\beta(1-\beta)}}arctanh[\frac{\beta(2r+r_0)}{r_0\sqrt{3\beta(4-\beta)}}][\alpha(\beta+1)+2\eta r_0^2]\\
&+\frac{\ln[(\beta-3)r_0^2+\beta r_0r
+\beta r^2]\{r_0^2(2\beta-3)[3\eta+(\alpha-2)\rho_0]-3\alpha\beta\}}{12\alpha\beta(1-\beta)}-\frac{1}{2}\ln(r),
\end{aligned}
\end{equation}
It is not difficult to prove that $\Phi(r)$ given by Eq. (66) is finite in the range (64). Thus, as the above-mentioned case, matching the interior spacetime geometry into an exterior vacuum geometry, this solution represents a traversable wormhole now. Comparing with the same case in \cite{45}, one can obtain a mathematical paradigm like Eq. (63) for the shape function. That means for different cosmological models, one can have the same shape function in the case of constant energy density.
\subsection{Isotropic pressure}
\begin{figure}
\centering
\includegraphics[scale=0.5]{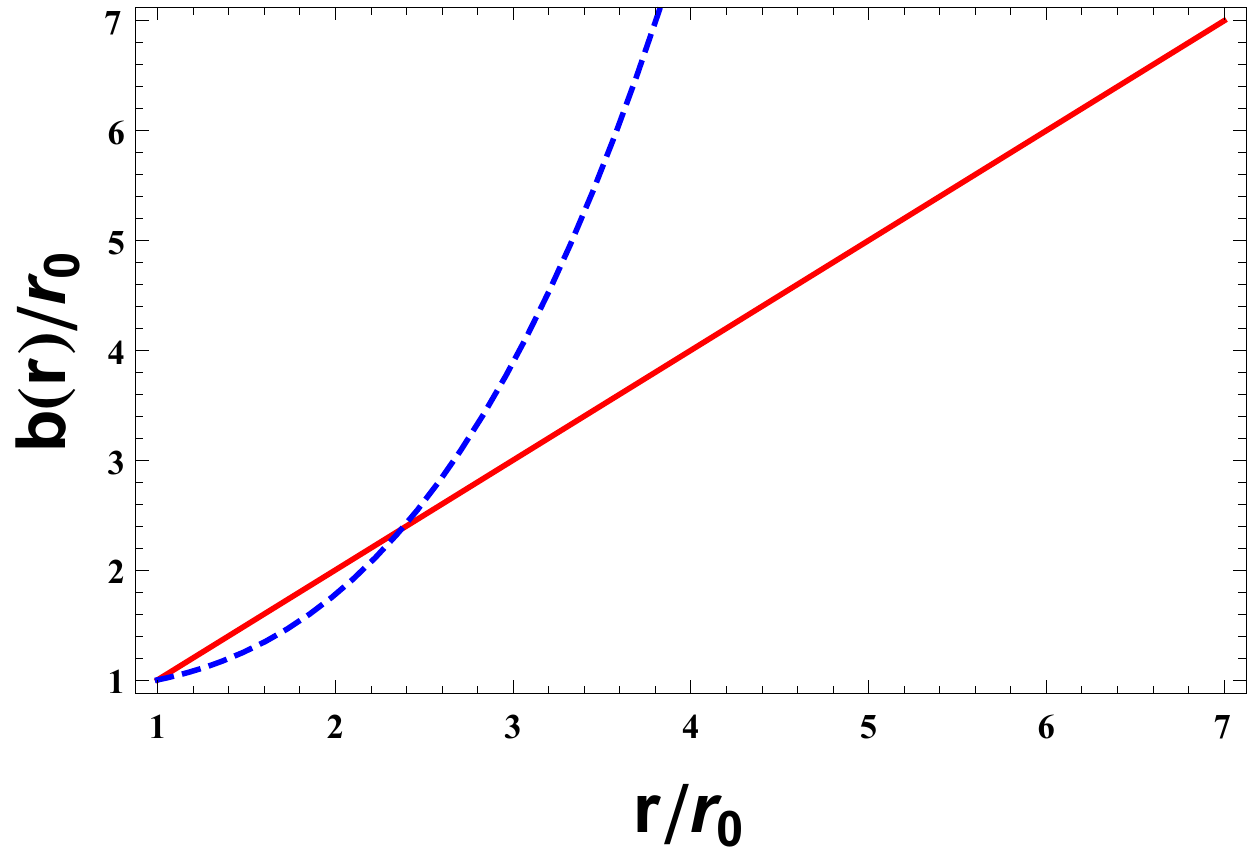}
\caption{To form a wormhole geometry, according to the restriction $b(r)<r$, one can find that the range below the solid (red) line represents  wormhole solutions. The solid (red) line corresponds to the line of $b(r)=r$, the dashed (blue) line corresponds to the line of $b(r)$ in the case of isotropic pressure (Here we take the values of the parameters $\eta\approx10^{-50}$ and $\alpha=0.423469 $). Moreover, one may discover that the dimensions of wormholes are substantially finite.}\label{fig7}
\end{figure}

Starting from Eq. (32) and considering an isotropic pressure, $p_r=p_t$, one can get the following differential equation:
\begin{equation}
\frac{(\frac{2}{3\alpha}-\frac{1}{3})\rho'}{\rho-(\frac{2}{3\alpha}-\frac{1}{3})(\rho-\frac{3\eta}{2-\alpha})}=\Phi'(r).
\end{equation}
It follows that
\begin{equation}
\rho(r)=\frac{C_1e^{\frac{2(2\alpha-1)\Phi(r)}{2-\alpha}}-3\eta}{2(2\alpha-1)}.
\end{equation}
Note that $\rho(r_0)=\frac{3\alpha(1+\eta r_0^2)}{r_0^2(2-\alpha)}$, we can obtain
\begin{equation}
C_1=\frac{3\alpha[2(2\alpha-1)+3\eta r_0^2]}{(2-\alpha)r_0^2}e^{\frac{2(2\alpha-1)\Phi(r_0)}{2-\alpha}}.
\end{equation}
Then, replacing Eq. (68) in Eq. (30) and solving it when taking the redshift function as $\Phi(r)=\ln(\frac{r}{r_0})$, one can get the shape function in the following manner:
\begin{equation}
b(r)=3r\{\frac{\eta}{2(1-2\alpha)}+\frac{\alpha r_0^{\frac{2(2\alpha-1)}{2-\alpha}}r^{\frac{6(1-\alpha)}{2-\alpha}}(4\alpha+3\eta r_0^2-2)}{2(1-2\alpha)(7\alpha-8)}\}
\end{equation}
It is easy to be checked that this solution satisfy the flaring out conditions $b'(r_0)\approx0.9<1$ and $b(r)<r$ when $r>r_0$, and is not asymptotically flat. Furthermore, as before, we make a plot to illustrate the dimensions of this wormhole is substantially finite (see Fig.~\ref{fig7}). It is worth pointing out one can still construct a traversable wormhole by matching the interior geometry to an exterior vacuum geometry. Therefore, the dimensions of the RDE wormhole in this case is not arbitrarily large, which is different from the case of constant redshift function.
\section{Discussions and conclusions}
Since the elegant discovery that our universe is undergoing an accelerated expansion, cosmologists have proposed many alternatives to explain the accelerated mechanism, which mainly include two classes: physical dark energy models and extended theories of gravity. Actually, one can find that the two classes of models are physically equivalent by rearranging the terms in the Einstein field equations. Up to now, we still do not determine what the nature of dark energy is. In this letter, we are very interested in exploring the wormhole physics of the RDE model, which is one of these popular dark energy models, by assuming the dark energy is distributed homogeneously in the whole spacetime.

In the past few years, there are a lot of papers investigating the wormholes spacetime and constraining the model parameters for various kinds of dark energy models. But, there is no one to study wormholes by using the astrophysical observations as the data support. In particular, we can discover, through the constraints of observations data-sets to some well known dark energy model, the evolution behavior of the equation of state parameter during the whole history of the universe could be well studied quantitatively. To be more precise, when the NEC is violated, namely, the equation of state parameter is less than $-1$, wormholes may appear (open).
Since we think the astrophysical observations contain the most realistic physics, so we introduce naturally the astrophysical observations into the wormhole research, which seems to be the first try in the literature. For an concrete instance, we explore the traversable wormholes in the RDE model.

In the present paper, we make a brief review on the RDE model and constrain this model by astrophysical observations. Subsequently, we find out the best fitting values of the parameters in this model by using the usual $\chi^2$ statistics, in order that we can know more about the evolution behavior of the universe and determine the evolution of the effective equation of state parameter $\omega_X$ with time. Then, one can investigate the RDE traversable wormholes better after an accurate equation of state is obtained. We have analyzed the effective equation of state of the RDE model and give the constraint relation of the parameters by using the flaring out conditions. Furthermore, we have investigated some specific solutions and the related physical properties and characteristics by considering three different redshift functions, one specific shape function, constant dark energy density and isotropic pressure. In the first case, we find that the traversable wormhole dimension is finite and calculate out the traversal velocity and traversal time derived from the so-called traversability conditions for a interstellar traveler. In the second one, we quantify the `` total amount '' of energy condition violating matter by computing the `` volume integral quantifier '', and find that one may theoretically construct a traversable wormhole with infinitesimal amounts of ANEC violating RDE matter. In the third case, we get a mathematical paradigm by  computing $b'(r_0)$ in the first three cases, and provide some interesting physical explanations. Actually, this paradigm also can be obtained from the violation of the NEC evaluated at the throat, namely, $p_r(r_0)+\rho(r_0)<0$. In the fourth case, we consider a specific shape function and obtain a traversable wormhole by matching the interior spacetime into an exterior vacuum spacetime. The fifth case also reflects a non-asymptotically flat spacetime, where the exotic matter from the RDE fluids is distributed in the vicinity of the throat. For the case of isotropic pressure, one may discover that the dimensions of wormholes is substantially finite.

After the general relativity's centennial, we are still confused with the attractive and mysterious nature of dark energy and dark matter in different scales, if we assume the dark sector is permeated everywhere in the whole universe. Wormholes are theoretically objects in the universe which now appear to attract more observational astrophysics interests and may provide a new window for new physics. In this situation, we apply the astrophysical data-sets into the wormhole physics and investigate six specific solutions quantitatively, which seems to be the first time in the wormhole research. Through astronomical observations, one can make a constraint on the parameters of a cosmological model, explore the type of cosmic matter in different stages of the universe, limit the number of available models for wormhole research, reduce the number of the wormholes corresponding to different parameters for a concrete cosmological model and provide a more clear picture for wormhole research from the new perspective of observational cosmology backgroud.

The future work could be to consider an obvious relation between the energy density and the transverse pressure, explore the profound connection between wormholes and energy conditions, and investigate the evolution of wormhole structure with time.
\section{acknowledgements}
Useful communications with Saibal Ray are highly appreciated over long time.
We thank Prof. Jing-Ling Chen for helpful discussions and comments, and Guang Yang and Sheng-Sen Lu for programming. This work is supported in part by the National Science Foundation of China.

\end{document}